\documentstyle{article}

\newtheorem{theorem}{Theorem}
\newtheorem{lemma}{Lemma}
\newtheorem{remark}{Remark}

\def\H{{\cal H}} % Hilbert space
\def\Hn{{\cal H}^{\otimes n}} % n tensor Hibert space
\def\X{{\cal X}} % input alphabet
\def\C{{\cal C}} % code book
\def\Pe{\mbox{\rm Pe}} % avarage error probability
\def\E{\mbox{\rm E}} % mean value
\def\Prob{\mbox{\rm Prob}} % probability
\def\P{{\cal P}} % set of probabilities
\def\Tr{\mbox{\rm Tr}\,} % trace
\def\Re{\mbox{\rm \bf R}} % Real
\def\diag{\mbox{\rm diag}} % diagonal matrix

\renewcommand{\epsilon}{\varepsilon}

\def\appendix{\par
    \setcounter{section}{0}\setcounter{subsection}{0}
    \def\thesection{\Alph{section}} \section*{Appendix}
}
\def\abstract{%
    \if@twocolumn
      \small\it Abstract\/\bf---$\!$%
    \else
      \begin{center}\vspace{-0.8em}\small\bf Abstract\end{center}\quotation\small
    \fi}
\def\endabstract{\vspace{0.6em}\par\if@twocolumn\else\endquotation\fi
    \normalsize\rm}

\def\keywords{\vspace{-.3em}
    \if@twocolumn
      \small\it Keywords\/\bf---$\!$%
    \else
      \begin{center}\small\bf Keywords\end{center}\quotation\small
    \fi}
\def\endkeywords{\vspace{0.6em}\par\if@twocolumn\else\endquotation\fi
    \normalsize\rm}

\begin{document}
%\bibliographystyle{IEEE}
%\nocite{*}%

\title{
Strong Converse to \\
the Quantum Channel Coding Theorem
}

\author{Tomohiro Ogawa and Hiroshi Nagaoka
\thanks{
The authors are with
the Graduate School of Information Systems,
The University of Electro-Communications,
1--5--1 Chofugaoka, Chofu, Tokyo 182--8585, Japan.
}
\thanks{
E-mail: ogawa@hn.is.uec.ac.jp, nagaoka@is.uec.ac.jp
}
}
\date{}

\maketitle

\begin{abstract}
 A lower bound on the probability of decoding error of quantum
communication channel is presented. The strong converse to the quantum
channel coding theorem is shown immediately from the lower bound.  It is
the same as Arimoto's method except for the difficulty due to
non-commutativity.
\end{abstract}


\section{Introduction}

Recently, the {\em quantum channel coding theorem} was established
by Holevo \cite{Holevo-QCTh}
and by Schumacher and Westmoreland \cite{Schumacher-Westmoreland},
after the breakthrough of
Hausladen {\it et al.} \cite{Hausladen-et-al.}.
Furthermore, a upper bound on the probability of
decoding error, in case rate below capacity, was derived
by Burnashev and Holevo \cite{Bur-Holevo}.
It is limited in pure signal state.
They conjectured on a upper bound in general signal state,
which corresponds to Gallager's bound \cite{Gallager} in
classical information theory.
We will show a lower bound on the probability of decoding error,
in case rate above capacity, 
which corresponds to Arimoto's bound \cite{Arimoto}.
The {\em strong converse} to the quantum channel coding theorem
is shown immediately from the lower bound.

Let $\H$ be Hilbert space which represents a physical system
of information carrier. We suppose $\dim \H < \infty$ for simplicity.
{\em Quantum channel} \cite{Holevo-channel} is defined as mapping 
$i\in \X \mapsto \rho_i \, (i=1,\cdots,a)$, where 
$\X = \{1,\cdots,a\}$ is the set of input
alphabet and $\rho_i\,(i=1,\cdots,a)$ is a {\em density operator}
in $\H$, i.e., non-negative operator with trace one.
For a more general treatment,
see Fujiwara and Nagaoka \cite{Fujiwara-Nagaoka}.

To describe asymptotic property, we use $n$-th extension of the channel.
The {\em messages} $\{1,\cdots,M_n\}$ is encoded to
a {\em codebook} $\C^{(n)}=\{ u^1,\cdots,u^{M_n} \}$, where
each $u^k = i_{1}^k\cdots i_{n}^k \in \X^n$ $(k=1,\cdots,M_n)$
is a {\em codeword}, and is mapped to
$\rho_{u^k}=\rho_{i_1^k} \otimes \cdots \otimes \rho_{i_n^k}$,
which is a density operator in $\Hn$.
Decoding process $X^{(n)}=\{X_0,X_1,\cdots,X_{M_n}\}$ 
is a {\em quantum measurement} \cite{Holevo-channel},
that is a resolution of identity in $\Hn$, i.e.,
$X_k \ge 0 \, (k=0,\cdots,M_n)$ and $\sum_{k=0}^{M_n} X_k =I$.
We think of $X_0$ as evasion of decoder.
A pair of encoding and decoding process $(\C^{(n)},X^{(n)})$
is called a {\em code} with cardinality $M_n$.
$R_n=\log M_n / n$ is called {\em transmission rate}
for a code $(\C^{(n)},X^{(n)})$.
In the sequel, we will omit the subscript $n$
when no confusion is likely to arise.

The conditional probability of output $k$, 
when message $l$ was sent, is given by
$P(k|l)=\Tr \rho_{u^l} X_k$.
If all messages arise with uniform probability,
the {\em average error probability} of code $(\C,X)$ is
\begin{eqnarray*}
\Pe(\C,X) = 1-\frac{1}{M}\sum_{k=1}^M \Tr \rho_{u^k} X_k
\end{eqnarray*}
Let us denote the minimum of the average error probability as
\begin{eqnarray*}
\Pe(M_n,n) = \min_{\C,X} \Pe(\C,X)
\end{eqnarray*}
The (operational) {\em capacity} \cite{Holevo-capacity}
is defined as the number $C$
such that $\Pe(e^{nR},n)$ tends to zero as $n\rightarrow \infty$
for any $0\le R<C$ and does not tend to zero if $R>C$.

Let $\pi=\{\pi_i\}_{i=1}^a$ be a probability distribution
on $\X$, and
define (formal) {\em quantum mutual information}
\cite{Holevo-channel} as
\begin{eqnarray*}
I(\pi) = 
H(\overline{\rho}_{\pi}) - \sum_{i=1}^a \pi_i H(\rho_i)
\end{eqnarray*}
where
$\overline{\rho}_{\pi} = \sum_{i=1}^a \pi_i \rho_i$
and $H(\rho)=-\Tr \rho \log \rho$, which is Von Neumann entropy.
The quantum coding theorem states that $\max_{\pi} I(\pi)$
is equal to the operational capacity $C$.
The aim of this correspondence is to show the strong converse
to the quantum channel coding theorem, i.e., 
$\Pe(e^{nR},n)$ tends to one exponentially
as $n \rightarrow \infty$ if $R>C$.

\section{lower bound on the average error probability}

To begin with, we will show the following Lemma.
\begin{lemma}
For an arbitrary measurement $X=\{X_k\}_{k=0}^{M}$
\begin{eqnarray}
\Pe(\C,X) \ge
 1- \frac{1}{M}
 \Tr \left(
      \sum_{k=1}^M \rho_{u^k}^{\frac{1}{\beta}}
      \right)^{\beta}
  \quad (0< \beta \le 1)
  \label{sitaosae}
\end{eqnarray}
holds.
\label{lemma1}
\end{lemma}
\begin{proof}
$
 \rho_{u^l}^{\frac{1}{\beta}}
\le
 \sum_{k=1}^M \rho_{u^k}^{\frac{1}{\beta}}
\; (l=1,\cdots,M)
$
is obvious.
Since $x^{\beta}\; (0 < \beta \le 1)$
is a {\em operator monotone function}
(see {\it ex.} \cite{Hansen-Pedersen}),
\begin{eqnarray}
\rho_{u^l} = \left(
	  \rho_{u^l}^{\frac{1}{\beta}}
	  \right)^{\beta}
\le
\left(
 \sum_{k=1}^M \rho_{u^k}^{\frac{1}{\beta}}
 \right)^{\beta}
\quad (l=1,\cdots,M)
\end{eqnarray}
holds. Hence,
\begin{eqnarray*}
 \Pe(\C,X) &=& 1-\frac{1}{M}\sum_{l=1}^M \Tr \rho_{u^l} X_l \\
 &\ge& 1-\frac{1}{M}\sum_{l=1}^M 
  \Tr \left(
       \sum_{k=1}^M \rho_{u^k}^{\frac{1}{\beta}}
       \right)^{\beta} X_l \\
 &\ge&  1- \frac{1}{M}
 \Tr \left(
      \sum_{k=1}^M \rho_{u^k}^{\frac{1}{\beta}}
      \right)^{\beta}
\end{eqnarray*}
where we used
$\sum_{l=1}^M X_l \le I$
in the last inequality.
\end{proof}

Following Arimoto \cite{Arimoto},
let us apply random coding technic to
Lemma \ref{lemma1} with a probability distribution
\begin{eqnarray*}
P(u^1,\cdots,u^M)  = \Prob\{\C = (u^1,\cdots,u^M)\}
\end{eqnarray*}
For this purpose, we shall need next two conditions.
\begin{enumerate}
 \item $\displaystyle{
       \E_{P} \left[ \min_{X} \Pe(\C,X) \right]
       =\min_{\C} \min_X \Pe(\C,X)
       }$
 \item $P(u^1,\cdots,u^M)$ is invariant
       under a permutation of $u^1,\cdots,u^M$.
\end{enumerate}
Actually, such a probability distribution
on the set of all codebook exists.
Suppose that $\;\hat{\C}=(\hat{u}^1,\cdots,\hat{u}^M)$
attains the minimum of condition 1, then from symmetry of average error
probability, a permutation of
$(\hat{u}^1,\cdots,\hat{u}^M)$
also attains the minimum. Therefore
\begin{eqnarray*}
\hat{P}(u^1,\cdots,u^M)=\left\{
  \begin{array}{ll}
   \frac{1}{M!} & \mbox{$(u^1,\cdots,u^M)$ is} \\
                & \mbox{a permutation of $(\hat{u}^1,\cdots,\hat{u}^M)$}\\
   0            & \mbox{otherwise}
  \end{array}
\right.
\end{eqnarray*}
is the probability distribution which satisfies above two conditions.
Furthermore, the marginal probability distributions
of $\hat{P}(u^1,\cdots,u^M)$
does not depend on $u^1,\cdots,u^M$ by condition 2, i.e.,
\begin{eqnarray*}
 && \hat{P}(u^1)= \cdots =\hat{P}(u^M) \\
 && = \sum_{u^2\in \X^n}\cdots\sum_{u^M\in \X^n}\hat{P}(u^1,u^2,\cdots,u^M)
\end{eqnarray*}
By taking average of (\ref{sitaosae}) with $\hat{P}$,
we obtain
\begin{eqnarray}
 && \min_{\C,X} \Pe(\C,X) 
 = %%%%%%%%%%%%%%%%%%%%%%%%%%%%%%%%%%%%%%%%%%%%
 \E_{\hat{P}} \left[ \min_{X} \Pe(\C,X) \right]
 \nonumber \\
 &\ge& %%%%%%%%%%%%%%%%%%%%%%%%%%%%%%%%%%%%%%%%%%%%
 1- \frac{1}{M} \E_{\hat{P}}
 \left[
  \Tr \left(
       \sum_{k=1}^M \rho_{u^k}^{ \frac{1}{\beta} }
       \right )^{\beta}
  \right]
   \label{beforeJensen}
\end{eqnarray}
Using Jensen type inequality, which is derived
from {\em operator concavity} of $x^{\beta}\;(0 < \beta \le 1)$
(see \cite{Hansen-Pedersen}),
(\ref{beforeJensen}) is bounded as
\begin{eqnarray}
 &\ge& %%%%%%%%%%%%%%%%%%%%%%%%%%%%%%%%%%%%%%%%%%%%
 1- \frac{1}{M}
 \Tr \left(
      \sum_{k=1}^M \E_{\hat{P}} \left[
		 \rho_{u^k}^{ \frac{1}{\beta} }
		 \right]
      \right)^{\beta} \label{Jensen}\\
 &=& %%%%%%%%%%%%%%%%%%%%%%%%%%%%%%%%%%%%%%%%%%%%
 1- M^{\beta -1}
 \Tr \left(
      \sum_{u \in \X^n} \hat{P}(u)
	  \rho_{u}^{ \frac{1}{\beta} }
      \right)^{\beta} \\
 &\ge& %%%%%%%%%%%%%%%%%%%%%%%%%%%%%%%%%%%%%%%%%%%%
 1- M^{\beta -1} \max_{P\in \P_{\X^n}}
 \Tr \left(
      \sum_{u \in \X^n} P(u)
	  \rho_{u}^{ \frac{1}{\beta} }
      \right)^{\beta}
  \label{max_Xn}
\end{eqnarray}
where we used the notation $\P_{\X^n}$, which represents
the set of all the probability distributions on $\X^n$.

The following Lemma is the same as classical information
theory (see \cite{Gallager}\cite{Arimoto})
except for one point that derivative
of a function is not easy due to non-commutativity. We will give the
proof for convenience.
\begin{lemma}
Let $\P_{\X}$ be the set of 
all the probability distributions on $\X$. Then
\begin{eqnarray*}
 \max_{P\in \P_{\X^n}}
  \Tr \left(
       \sum_{u \in \X^n} P(u)
       \rho_{u}^{ \frac{1}{\beta} }
       \right)^{\beta}
   =
   \left[
    \max_{\pi \in \P_{\X}}
    \Tr \left(
	 \sum_{i=1}^a \pi_i \rho_i^{ \frac{1}{\beta} }
	 \right)^{\beta}
    \right]^n
\end{eqnarray*}
\label{maxlemma}
\end{lemma}
\begin{proof}
Let us define a function of $\pi \in \P_{\X}$ as
\begin{eqnarray*}
 f(\pi) = \Tr \left(
	     \sum_{i=1}^a \pi_i \rho_i^{ \frac{1}{\beta} }
	     \right)^{\beta}
\end{eqnarray*}
We note that $f$ is a concave function.
First, we show that necessary and sufficient condition on
the probability distribution $\pi^* \in \P_{\X}$,
which attains the maximum of $f(\pi)$, is
\begin{eqnarray}
 && \Tr S^{\beta-1} \rho_i^{\frac{1}{\beta}} \le \Tr S^{\beta}
  \quad \mbox{(equality if $\pi_i^* > 0$)}
  \label{KTcond1}\\
 && \mbox{where} \quad
  S =  \sum_{i=1}^a \pi_i^* \rho_i^{ \frac{1}{\beta} }
  \label{KTcond2}
\end{eqnarray}
To show this, introduce Lagrange multiplier
$s_i\;(s_i\ge 0,i=1,\cdots,a)$ and $\lambda$,
define another function of $\pi$ as
\begin{eqnarray*}
 g(\pi) =
  - f(\pi) - \sum_{i=1}^a \pi_i s_i -\lambda (\sum_{i=1}^a \pi_i -1)
\end{eqnarray*}
differentiate $g(\pi)$ by $\pi_i$ and make it to $0$.
From general theory of Lagrange multiplier method
(see {\it ex.} \cite{Luenberger}),
we assert that necessary and sufficient condition on
$\pi^*\in \P_{\X}$, which attains maximum of $f(\pi)$
(i.e. minimum of $g(\pi)$),
is that there exist $s_i\ge 0\;(i=0,\cdots,a)$ and $\lambda$
which satisfies next conditions.
\begin{eqnarray}
-\beta \; \Tr S^{\beta-1} \rho_i^{\frac{1}{\beta}}
 - s_i - \lambda =0
 \label{lagbibun}\\
\sum_{i=1}^a \pi_i^* s_i =0
 \label{spi}
\end{eqnarray}
where we used derivative of $f(\pi)$ 
(see Appendix Lemma \ref{lemma:derivative})
\begin{eqnarray*}
 \frac{\partial f}{\partial \pi_i} =
  \Tr \left(
       \sum_{i=1}^a \pi_i \rho _i^{\frac{1}{\beta}}
       \right)^{\beta-1} \rho _i^{\frac{1}{\beta}}
\end{eqnarray*}
By multiplying $\pi_i^*$ to the both sides of (\ref{lagbibun})
and summing over, we obtain
\begin{eqnarray*}
-\lambda = \beta \; \Tr S^{\beta} 
\end{eqnarray*}
Meanwhile, $s_i=0$ if $\pi_i^*>0$ by (\ref{spi}).
Hence (\ref{lagbibun})(\ref{spi}) is equivalent to
\begin{eqnarray*}
 \exists s_i \ge 0  \quad \mbox{s.t.} \quad
 \Tr S^{\beta-1} \rho_i^{\frac{1}{\beta}} =
  - \frac{s_i}{\beta} + \Tr S^{\beta} \\
 \mbox{($s_i=0$ if $\pi_i^*>0$)}
\end{eqnarray*}
Moreover, this is equivalent to (\ref{KTcond1})(\ref{KTcond2}).
Now, Suppose $\pi^*$ satisfies (\ref{KTcond1})(\ref{KTcond2}),
and put
\begin{eqnarray*}
P^*(u)=P^*(i_1,\cdots,i_n)=
 \pi_{i_1}^* \cdots \pi_{i_n}^*
\end{eqnarray*}
which is i.i.d. extension of $\pi^*$.
Then it is clear $P^*$ satisfies
\begin{eqnarray*}
 && \Tr {\tilde{S}}^{\beta-1} \rho_u^{\frac{1}{\beta}}
  \le \Tr {\tilde{S}}^{\beta}
  \quad \mbox{(equality if $P^*(u) > 0$)} \\
 && \mbox{where} \quad
  {\tilde{S}} =  \sum_{u\in \X^n} P^*(u) \rho_u^{ \frac{1}{\beta} }
\end{eqnarray*}
Hence
\begin{eqnarray*}
 && \max_{P\in \P_{\X^n}}
  \Tr \left(
       \sum_{u \in \X^n} P(u) \rho_{u}^{ \frac{1}{\beta} }
       \right)^{\beta} \\
 &=&
  \Tr \left(
       \sum_{i_1}\cdots\sum_{i_n}  \pi_{i_1}^* \cdots \pi_{i_n}^*
       (
       \rho_{i_1} \otimes \cdots \otimes \rho_{i_n}
       )^{ \frac{1}{\beta} }
       \right)^{\beta} \\
 &=&
   \left[
    \Tr \left(
	 \sum_{i=1}^a \pi_i^* \rho_i^{ \frac{1}{\beta} }
	 \right)^{\beta}
    \right]^n \\
 &=&
   \left[
    \max_{\pi \in \P_{\X}}
    \Tr \left(
	 \sum_{i=1}^a \pi_i \rho_i^{ \frac{1}{\beta} }
	 \right)^{\beta}
    \right]^n
\end{eqnarray*}
\end{proof}

Now, from (\ref{max_Xn}) and Lemma \ref{maxlemma} we obtain
\begin{eqnarray*}
 && \min_{\C,X} \Pe(\C,X) \\
 &\ge& %%%%%%%%%%%%%%%%%%%%%%%%%%%%%%%%%%%%%%%%%%%%
 1- M^{\beta -1}
 \left[
  \max_{\pi \in \P_{\X}} \Tr \left(
       \sum_{i=1}^a \pi_i \rho_i^{ \frac{1}{\beta} }
       \right)^{\beta}
  \right]^n
\end{eqnarray*}
Let us put $s=\beta-1$, recall $R=\log M/n$ and define
\begin{eqnarray}
 E_0(s,\pi)=
  - \log \left(
	  \Tr \left(
	       \sum_{i=1}^a \pi_i \rho_i^{ \frac{1}{s+1} }
	       \right)^{s+1}
	  \right) \label{E(s,pi)}
\end{eqnarray}
then we have proved the following theorem.
\begin{theorem}
For all code $(\C,X)$
\begin{eqnarray}
 && \Pe(\C,X) \nonumber \\
 &\ge&
  1- \exp
  \left[
   -n \left[
       - s R + \min_{\pi \in \P_{\X}} E_0( s,\pi )
       \right]
   \right]  \label{rand_low} \\
  && ( -1 < s \le 0 ) \nonumber
\end{eqnarray}
\end{theorem}
\begin{remark}
(\ref{E(s,pi)}) has appeared in \cite{Bur-Holevo}
as a conjecture on the upper bound on the average error probability,
which forms dual with (\ref{rand_low}).
They proved it in case that all $\rho_i\,(i=1,\cdots,a)$ are pure.
\end{remark}

\section{Strong converse to the quantum channel coding theorem}

To understand the graph of $E_0(s,\pi)$,
we show the following lemma.
\begin{lemma}
\begin{eqnarray}
 && E_0(0,\pi)=0
  \label{E1}\\
 && E_0(s,\pi) \le 0,\quad (-1 < s \le 0)
    \label{E2}\\
 && \frac{\partial E_0(s,\pi)}{\partial s} \ge 0,
  \quad (-1 < s \le 0)
    \label{E3}\\
 && \left. \frac{\partial E_0(s,\pi)}{\partial s} \right|_{s=0}
  = I(\pi)
  \label{E4}
\end{eqnarray}
\label{lemma:graph}
\end{lemma}
\begin{proof}
(\ref{E1}) is obvious.
From the following Lemma, $E_0(s,\pi)$ is shown to be non-decreasing
in $(-1,0\,]$. That is why (\ref{E2})(\ref{E3}) holds.
Using Appendix Lemma \ref{lemma:derivative}, 
we can calculate directly
the derivative of $E_0(s,\pi)$ to obtain (\ref{E4}),
or using
\begin{eqnarray*}
&& \left.
 \frac{\partial}{\partial s} \Tr \left(
   \sum_{i=1}^a \pi_i \rho_i^{ \frac{1}{s+1} }
 \right)^{s+1} 
\right|_{s=0} \\
&=&
\left.
 \frac{\partial}{\partial s} f(s,t)
\right|_{{}^{s=0}_{t=0}}
+
\left.
 \frac{\partial}{\partial t} f(s,t)
\right|_{{}^{s=0}_{t=0}} \\
&=&
 \sum_{i=1}^a \pi_i H(\rho_i) - H(\overline{\rho}_{\pi})
\end{eqnarray*}
where we put
$
f(s,t)=
\Tr \left( \sum_{i=1}^a \pi_i \rho_i^{ \frac{1}{s+1} } \right)^{t+1}
$ ,
we obtain
\begin{eqnarray*}
\left. \frac{\partial E_0(s,\pi)}{\partial s} \right|_{s=0}
&=&
-\left.
  \frac{
    \frac{\partial}{\partial s} \Tr \left(
      \sum_{i=1}^a \pi_i \rho_i^{ \frac{1}{s+1} }
    \right)^{s+1}
  }{
    \Tr \left(
      \sum_{i=1}^a \pi_i \rho_i^{ \frac{1}{s+1} }
    \right)^{s+1}
  }
\right|_{s=0} \\
&=& I(\pi)
\end{eqnarray*}
\end{proof}

\begin{remark}
Burnashev and Holevo \cite{Bur-Holevo} showed,
in case that all $\rho_i\,(i=1,\cdots,a)$ are pure and $0\le s\le 1$,
(\ref{E1})--(\ref{E4}) and
\begin{eqnarray*}
 \frac{\partial^2 E_0(s,\pi)}{\partial s^2} \le 0,
  \quad (0 \le s \le 1)
  \label{E5}
\end{eqnarray*}
\end{remark}

\begin{lemma}
\footnote{
We don't know where to refer about this lemma, but it is 
in \cite{Hiai-Yanagi} as a exercise.
}
Let 
$A_i \,(i=1,\cdots,a) $ be non-negative bounded operators in $\H$.
If $0 < \alpha \le \beta \le 1$
then
\begin{eqnarray}
\left(
 \sum_{i=1}^a \pi_i A_i^{\frac{1}{\alpha}}
 \right)^{\alpha}
\ge
\left(
 \sum_{i=1}^a \pi_i A_i^{\frac{1}{\beta}}
 \right)^{\beta}
\end{eqnarray}
\label{tantyouzouka}
\end{lemma}
\begin{proof}
First, we will show
\begin{eqnarray}
\sum_{i=1}^a \pi_i A_i \le
 \left(
  \sum_{i=1}^a \pi_i A_i^{\frac{1}{\gamma}}
  \right)^{\gamma}
  \label{gammaineq}
\end{eqnarray}
for $0<\gamma \le 1$ and $A_i \ge 0\;(i=1,\cdots,a)$.
Let us put unitary operator $U$
in $\H \oplus \cdots \oplus \H$ as
\begin{eqnarray}
U = \left(
   \begin{array}{ccc}
    \sqrt{\pi_1}I & & \\
    \vdots & * & \\
    \sqrt{\pi_a}I & &
   \end{array}
   \right)
\end{eqnarray}
and projection $P$ as
\begin{eqnarray}
P = \diag \left[I,0,\cdots,0\right]
\end{eqnarray}
Generally, for $0 < \gamma \le 1$ and 
operator $A$, $C$ $(||C|| \le 1)$,
$C^* A^{\gamma}C \le (C^*AC)^{\gamma}$
holds. (see \cite{Hansen-Pedersen}, or \cite{Oya-Petz}, p.18)
Using this property, we obtain
\begin{eqnarray*}
&&\diag
\left[
\sum_{i=1}^a \pi_i A_i \, , 0, \cdots, 0
\right] \\
&=& PU^*
\diag
\left[
A_1, \cdots, A_a
\right]
UP \\
&=& P U^*
\left(
 \diag
 \left[
  A_1^{\frac{1}{\gamma}}, \cdots, A_a^{\frac{1}{\gamma}}
  \right]
\right)^{\gamma}
U P \\
&=& P
\left(
 U^*
 \diag
 \left[
  A_1^{\frac{1}{\gamma}}, \cdots, A_1^{\frac{1}{\gamma}}
  \right]
  U
\right)^{\gamma}
P \\
&\le&
\left(
PU^*
\diag
\left[
  A_1^{\frac{1}{\gamma}}, \cdots, A_1^{\frac{1}{\gamma}}
\right]
UP
\right)^{\gamma} \\
&=&
\diag
\left[
\left(
  \sum_{i=1}^a \pi_i A_i^{\frac{1}{\gamma}}
 \right)^{\gamma},0,\cdots,0
\right]
\end{eqnarray*}
which shows (\ref{gammaineq}).
Now, change $\gamma$ into $\frac{\alpha}{\beta}$ and 
$A_i$ into $A_i^{\frac{1}{\beta}}$ in (\ref{gammaineq}),
then
\begin{eqnarray*}
 \sum_{i=1}^a \pi_i A_i^{\frac{1}{\beta}}
\le
\left(
 \sum_{i=1}^a \pi_i A_i^{\frac{1}{\alpha}}
 \right)^{\frac{\alpha}{\beta}}
\end{eqnarray*}
Since $x^{\beta}$ is a operator monotone function,
\begin{eqnarray*}
\left(
 \sum_{i=1}^a \pi_i A_i^{\frac{1}{\beta}}
 \right)^{\beta}
\le
\left(
 \sum_{i=1}^a \pi_i A_i^{\frac{1}{\alpha}}
 \right)^{\alpha}
\end{eqnarray*}
\end{proof}

Now, from Lemma \ref{lemma:graph} if $R>C$
there exist $-1<t<0$ such that
for all $s \in (t,0),\;-sR + \min_{\pi} E_0(s,\pi)>0$.
Thus, we state the following strong converse theorem.

\begin{theorem}
If $R>C$, then for all code $(\C,X)$,
$\Pe(\C,X)$ goes to 1 exponentially
as $n\rightarrow \infty$.
\end{theorem}

\section*{Acknowledgment}
The authors wish to thank Dr.~Keiji~Matsumoto for useful discussion
about Lemma \ref{lemma:derivative}.

\appendix

\begin{lemma}
Let $f: (a,b) \rightarrow \Re$ be an analytic function and
$X(t)$ a Hermitian with real parameter $t$,
the spectrum of which is in $(a,b)$.
Then,
\begin{eqnarray*}
\frac{\partial}{\partial t} \Tr f(X(t))
 = \Tr f'(X(t)) \frac{\partial X(t)}{\partial t}
\end{eqnarray*}
where $f'$ is the derivative of $f$.
\label{lemma:derivative}
\end{lemma}
\begin{proof}
Leu us expand $f$ around $x_0 \in (a,b)$ as
$f(x)=\sum_{n=0}^{\infty} a_n (x-x_0)^n$ 
\def\Xhat{\hat{X}}
and put $\Xhat(t)=X(t)-x_0I$.
We can calculate as follows.
\begin{eqnarray*}
&& \frac{\partial}{\partial t} \Tr f(X(t))
=
\sum_{n=0}^{\infty} a_n \Tr \frac{\partial}{\partial t} 
\Xhat(t)^n \\
&=&
\sum_{n=1}^{\infty} a_n \sum_{i=1}^n \Tr 
\biggl(
\Xhat(t) \cdots \Xhat(t) 
\underbrace{
\frac{\partial \Xhat(t)}{\partial t} 
}_{\mbox{$i$-th}}
\Xhat(t) \cdots \Xhat(t)
\biggr) \\
&=& 
\sum_{n=1}^{\infty} n a_n 
\Tr 
\biggl(
\Xhat(t)^{n-1} \frac{\partial}{\partial t} \Xhat(t)
\biggr) \\
&=& 
\Tr f'(X(t)) \frac{\partial X(t)}{\partial t}
\end{eqnarray*}
\end{proof}

%\bibliography{strong}

\end{document}